\documentclass[12pt,preprint]{aastex}

\shorttitle{Unusual Eclipse}
\shortauthors{Cohen, Herbst \& Williams}

\begin{document}

\title{An Unusual Eclipse of a Pre-Main Sequence Star in IC 348}

\author{Roger E. Cohen, William Herbst }
\and
\author{Eric C. Williams}
\affil{Astronomy Department, Wesleyan University, Middletown, CT 06459}

\begin{abstract}
A solar-like pre-main sequence star (TJ 108 = H 187 = LRLL 35 = HMW 15) in the extremely young cluster IC 348 has been found, which apparently experienced an eclipse lasting $\sim$3.5 years, much longer than has ever been detected for any normal eclipsing binary. The light curve is flat-bottomed and rather symmetric, with a depth of 0.66 mag in Cousins I. During eclipse, the system reddened by $\sim$0.17 mag in R-I.  We argue that the eclipsing body is not a star because of the small probability of detecting an eclipse in what would be a very widely separated binary. Instead, it appears that the eclipse was caused by a circumstellar or circumbinary cloud or disk feature which occulted the star, or one of its components, if it is a binary system. We emphasize the importance of more detailed study of this object, which appears to be a new member of a small class of pre-main sequence stars whose variability can be firmly linked to occultation by circumstellar (or circumbinary) matter.
\end{abstract}


\section{Introduction}

The importance of eclipses in astronomy is well known and includes the opportunity for accurate determination of fundamental stellar parameters, such as masses and radii, as well as the chance to study stellar features (e.g. spots) or circumstellar structures (e.g. gas streams) with high spatial resolution.  Unfortunately, very few pre-main sequence (PMS) stars are known to undergo eclipses, perhaps because the requisite edge-on geometry usually results in their being obscured by a circumbinary disk.  Exceptions include a few binary stars that are not very close to the birth line, EK Cep \citep{p87}, TY CrA \citep{c98} and RXJ 0529.4+0041 \citep{c00}  as well as the unusual, weak-lined T Tauri star (WTTS) KH 15D \citep{{h01},{h02}}, which is periodically eclipsed by circumstellar matter. \citet{s02} have very recently discovered one or more additional interesting eclipsing systems in Orion. Some authors argue that the characteristic irregular variability of a class of PMS objects known as UX Orionis stars is also indicative of episodic eclipse events \citep{g98}, although other interpretations are possible \citep{hs98}. Finally, it has recently been proposed that the largely irregular variations of the classical T Tauri star (CTTS) AA Tau may be understood as a result of eclipses by circumstellar matter \citep{b99}. 

Part of the difficulty in finding, or recognizing, eclipses among PMS systems is that almost all such stars are variable on time scales of hours to days due to the rotation and evolution of their spotted surfaces. Cool spots on WTTS and CTTS, analogous to sunspots but much larger, and hot spots or zones on CTTS due to magnetically channeled accretion are thought to be the principal causes of surface structure \citep{h94}. Photometric monitoring of PMS stars in nearby young clusters has been carried out at Van Vleck Observatory, on the campus of Wesleyan University, since the mid-1980Õs in an effort to elucidate these phenomena and to study stellar rotation. For the past five years, the well-known young cluster IC 348 \citep{{tj}, {h98}, {lrll}, {l03}} has been among the targets. Results from the first year of that monitoring effort were presented by \citet{hmw} and discussion of the full data set is in preparation \citep{chw}. Here we report on one remarkable star.

Analysis of the data during the last four years has revealed a star, TJ 108 = H 187 = LRLL 35 = HMW 15, whose light curve has unique and surprising properties. It appears to be an eclipse, but the duration of the event, about 3.5 years, greatly exceeds that of any known eclipsing system. (The current record for eclipse duration of $\sim$2 yr is held by $\epsilon$ Aur, a supergiant that  \citet{l96} model as being eclipsed by a circumstellar disk surrounding its binary companion.) Because of the unusual nature of this event and its possible importance for studies of star and planet formation, we describe its characteristics in this {\it Letter}. Although the current eclipse has ended, it is likely that additional ones will be seen in this star or others similar to it. Such events may be our best current method for finding and characterizing small scale structure in the circumstellar disks of solar-like PMS stars. 

\section{Observations and Results}

The photometric monitoring was carried out mostly by undergraduates using a CCD camera attached to the 0.6m Perkin telescope of the Van Vleck Observatory on the campus of Wesleyan University. Typically, we obtained five 1-minute exposures of each field through a Cousins I filter; these were combined into a single image, and flat-fielded with twilight flats. Differential aperture photometry was done with respect to comparison stars on the same frames. Details of these observations and their reduction will be presented by  \citet{chw}. Here we note that the differential magnitudes have a precision of $\sim$0.01 or better. We have transformed the instrumental magnitudes to the standard Cousins system using the procedure outlined by \citet{hmw}.  The adopted transformation between instrumental (i) and standard (I) magnitude was, 
$$I = i + 0.099 (R-I) + 12.074$$
A data table will be included in the second paper \citep{chw}. 

\begin{figure}
\plotone{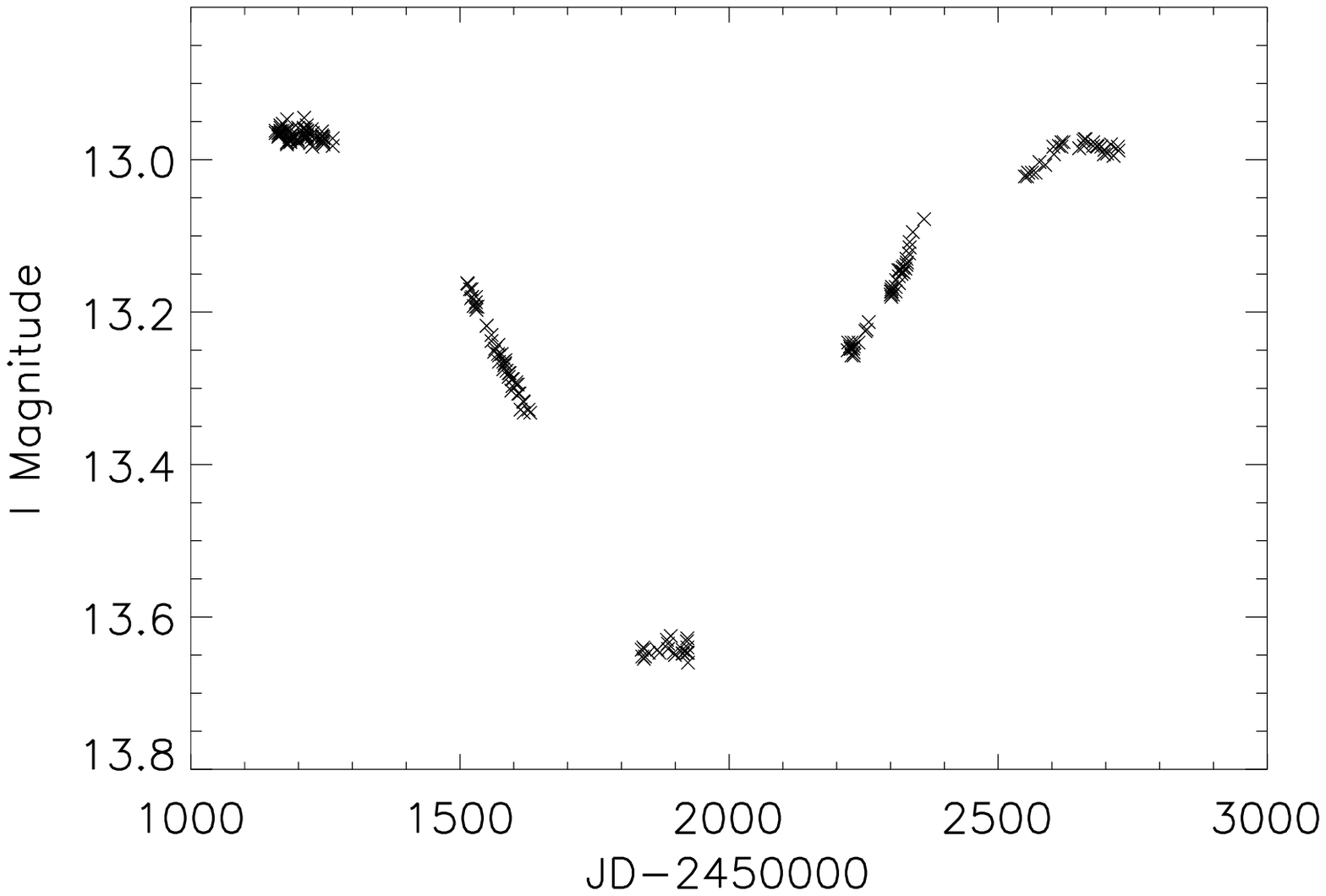}
\caption{The light curve of HMW 15 based on data obtained at Wesleyan.} 
\label{fig1}
\end{figure}

	The light curve for HMW 15 over five seasons is shown in Fig. \ref{fig1}. It has the general shape exhibited by an eclipsing binary. In particular, one sees a relatively stable star in the first season, which then fades steadily towards minimum light during the second season, shows a flat bottom to the eclipse in the third season, recovers steadily on approximately the same slope in season four and levels off close to its pre-eclipse brightness during the fifth season. A light curve like this would be unremarkable for an eclipsing binary, except that the time scale is astoundingly long --- much longer than has ever been seen in any such system. 

\begin{figure}
\plotone{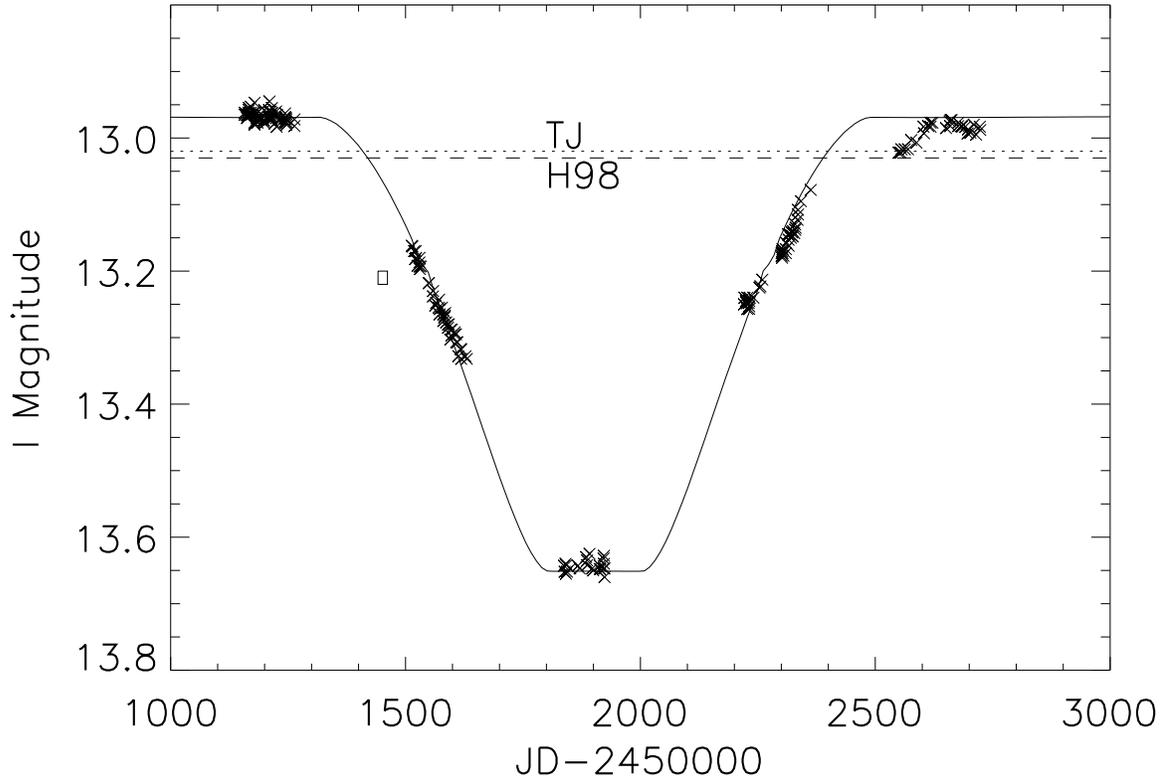}
\caption{A binary model for HMW 15 based on the parameters given in Table \ref{tab1}. Also shown are the published photometry of \citet{tj} and \citet{h98} based on measurements made from 1992 to 1996. See text for further discussion. The box indicates an observation reported by \citet{l03}.}
\label{fig2}
\end{figure}

	Available information on HMW 15 from the literature is summarized in Table \ref{tab1}. It appears to be an ordinary late-G or early-K PMS member of IC 348, which has an age of $\sim$2 Myr \citep{l03}.  Previously published photometry placed it at I=13.02, but \citet{h98} noted the star as a variable and he and C. Jordi kindly revisited their photometry for us. Herbig measured the star on 22 Nov 1993, 29 Nov 1994 and 21 Oct 1996, to be I = 12.97,13.07 and 13.57, respectively, while Trullols \& Jordi measured it on 24 Oct 92 and 9 - 11 Dec. 94 to be I = 13.36 and 12.94, respectively. The 1993 and 1994 measurements are consistent with our out-of-eclipse data, especially considering that at least a small amount of variability is expected for a PMS star of this type due to the usual spottedness of weak-lined T Tauri stars. The Oct 1992 and Oct 1996 data, on the other hand, suggest that the eclipse could be recurrent with a period as short as four years, or that there are additional occultation events (unrelated to the eclipse) affecting the light of this star.  One I-band measurement taken during ingress is reported in the literature (Luhman et al. 2003); they find I=13.21, which is roughly consistent with our light curve at that phase (see Fig. \ref{fig2}). 
	
\clearpage
\begin{deluxetable}{ll}
\scriptsize     
\tablecaption{Properties of HMW 15 \label{tab1}}
\tablewidth{0pt}
\tablehead{\colhead{Identifier} & \colhead{Source}}\startdata 
TJ 108 & Trullols \& Jordi (1997) \\
H 187 & Herbig (1998) \\
HMW 15 & Herbst, Maley \& Williams (2000) \\
LRLL 35 & Luhman et al. (1998), Luhman et al. (2003) \\
CXOPZ 153 & Preibsch \& Zinnecker (2001) \\
CXOPZ J034439.2+320736 & Chandra X-ray source \\
\hline
\multicolumn{2}{c}{Coordinates}\\
\hline
ICRS 2000.0 & 03 44 39.31 +32 07 34.0 \\
\hline
\multicolumn{2}{c}{Colors}\\
\hline
\multicolumn{2}{l}{V=15.82 (TJ), 15.90 (H98)}\\
\multicolumn{2}{l}{B-V=2.08 (TJ), 2.19 (H98)}\\
\multicolumn{2}{l}{U-B=1.48 (TJ)}\\
\multicolumn{2}{l}{V-I=2.80 (TJ), 2.87 (H98)}\\
\multicolumn{2}{l}{V-R=1.47 (TJ), 1.47 (H98)}\\
\multicolumn{2}{l}{J-H=0.96 (LRLL), 0.88 (2MASS)}\\
\multicolumn{2}{l}{H-K=0.40 (LRLL), 0.40 (2MASS)}\\
\multicolumn{2}{l}{K=9.90 (LRLL), 9.55 (2MASS)}\\
\hline
\multicolumn{2}{l}{Spectral Type: G8 (H98), G8-K4 or K3-K6 (IR, LRLL)}\\
\multicolumn{2}{l}{T$_{e}$=4730K (LRLL)}\\
\multicolumn{2}{l}{L$_{bol}$=2.07L$_{\sun}$ (LRLL)}\\
\multicolumn{2}{l}{R=2.15R$_{\sun}$ (based on T$_{e}$ and L$_{bol}$)}\\
\hline \hline
\multicolumn{2}{c}{Binary Model Parameters}\\
\hline 
T$_{1}$, T$_{2}$ & 5570K, 4500K\\
inclination,  R$_{1}$& 89.425$^{\rm{o}}$, 0.625R$_{2}$ \\
\enddata
\end{deluxetable}

\clearpage

	Mean colors from \citet{tj} and \citet{h98} are given in Table \ref{tab1} and were based on data obtained in 1993 and 1994, i.e. at maximum light; the out-of-eclipse star has an R-I value of 1.40. We normally observe only in I at VVO, but purely by accident obtained measurements in the R filter on many nights at the end of the third season, while the star was at minimum light. We found a color at minimum light for the star of R-I = 1.57 mag, significantly redder than at maximum. Herbig (private communication) also measured the star to be redder in both B-V (by 0.06 mag) and V-I (by 0.16 mag) during its faint excursion in 1996. It seems clear, therefore, that the star does get redder as it fades, although not as rapidly as one would expect based on an interstellar extinction law (see below). 
	
	A range of spectral types (from G8 to K3-K6) has been reported for the star, as summarized in Table \ref{tab1}. Whether this is simply observational scatter or not is unclear. There are other stars in IC 348 with similarly large ranges in spectral type \citep{l03} and this may simply reflect the uncertainties of the procedure, especially since different wavelength regions are used by different investigators. Hydrogen lines are in absorption and infrared data \citep{ll95} show no evidence for a circumstellar disk around the star. An active optics search for close binary companions by \citet{dbs} revealed no evidence for a companion within $\sim$0.25\arcsec, corresponding to about 80 AU at the distance of IC 348. Except for its unique light curve, HMW 15 appears to be an ordinary, solar-like PMS member of this extremely young cluster \citep{l03}. 

\section{Discussion}

We consider three possible scenarios to account for the unusual light curve: 1) an ordinary eclipsing binary composed of two PMS stars, 2) a single PMS star eclipsed by a circumstellar or intracluster dust cloud, or feature in a circumstellar disk and 3) a binary system in which one of the stars is eclipsed by an optically thick circumstellar, circumbinary or interstellar cloud or disk feature. While it is not possible to reach a definitive conclusion with the evidence currently available our analysis does lead us to favor the third hypothesis and we propose an observational test. 

\subsection {An eclipsing binary model} 
	It is possible to model the observed light curve with a standard eclipsing binary code as included, for example, in the program Binary Maker 2.0 \citep{b92}. A sophisticated approach is not needed here because effects such as reflection and distortion are negligible. In Fig. \ref{fig2}, we show the best fitting model light curve that could be found, by trial and error;  properties of this model are given in Table \ref{tab1}. We have been able to fit rather precisely not only the basic light curve shape, but also the color change at minimum light. Despite its success in explaining the light curve, there is actually little chance that this binary star model could be correct, for the following reason. The orbital velocity required to explain such a long duration for the eclipse in this scenario would need to be extremely small, namely  $\sim$0.05 km/s. Taking the masses of the components to be $\sim$1 M$_\odot$ this in turn would indicate a period for the orbit of about 200 Myr and a separation of $\sim$350,000 AU. Only one observer in $\sim$10$^{12}$ would be in the proper location to ever see an eclipse of this model system and it would only be seen during one part in 10$^8$ of each cycle. Roughly speaking, we would have had to monitor $\sim$10$^{17}$ more stars than we did to have a reasonable chance of seeing such an eclipse.    
	
\subsection{Eclipse of a single star by a circumstellar or intracluster cloud}
	It seems clear that, in this case, the eclipsing object must be much more extended than a star. It must be a circumstellar or circumbinary cloud or disk feature, or an interstellar cloud, possibly itself within the cluster. This would naturally account for the fact that a discovery like this is made in a star forming region, where such clouds are known to be ubiquitous. Since there is no evidence that HMW 15 is a binary star, we consider first that its fading is caused by a cloud or disk feature crossing our line of sight to a single star. Since the star does not completely disappear, we can conclude immediately that the occulting matter is either optically thin or covers only a portion of the star. The fact that the system reddens as it fades allows us to eliminate a purely optical thick model, which would predict little, if any, color change.
	
	In the optically thin model, the reddening observed during eclipse must be caused by selective extinction, which is what one would naturally expect from dust. We can compare the color change during eclipse to expectation based on an interstellar reddening law. Our data indicate
	$$ {\Delta (R - I) \over \Delta I} = {0.17 \over 0.66} = 0.26 \pm 0.04 $$
for the ratio of the color to magnitude change during eclipse. This may be compared with expectation based on a normal interstellar extinction law, E(R-I)/A$_I$= 0.39. At about the 3$\sigma$ level, we find that the color change observed is smaller than would be expected if the eclipse were caused by dust with a normal extinction law. This may mean that the dust grains are typically larger than those in the interstellar medium (perhaps implying grain growth in a circumstellar disk) or that the model of eclipse of a single star by an optically thin dust concentration is incorrect. It is interesting to note that Herbig's (private communication) single observation of the star in a faint state also yields a much smaller color change than would be expected if one were observing the effects of a passing cloud of normal interstellar dust. He found $\Delta$V/ $\Delta$(B-V) = 11.7, rather than the usual 3.1. The flat bottom and symmetry of the light curve also argue against a partially transparent cloud model and lead us to propose an alternative, which is something of a hybrid.
	
\subsection{Eclipse of one component of a binary star by an optically thick cloud}
	Suppose that HMW 15 is indeed a binary star and that one component is eclipsed, not by its companion star, but by an essentially opaque circumstellar, circumbinary or interstellar cloud or disk feature. In this case, we would expect to see a light curve nearly identical to that modeled in Fig. 2 and a color change consistent with our observations. We propose this scenario as, perhaps, the most attractive way to understand all of the currently available data. While the precise shape of the edge of the cloud or disk feature and its opacity structure would define the light curve in detail, it will approximately follow the curve defined by the binary model assuming that the feature is quite large and rather sharp-edged. As Fig. 2 shows, this is an excellent fit to our I photometry and the color change during eclipse is explained by the disappearance of light from the bluer component. 
	
	The basic geometry of this class of models is similar to what has been proposed for KH 15D by \citet{h02}. A sharp-edged, opaque cloud gradually occults a star due to their relative motions. From the light curve, we can measure t$_{in}$, the time for the cloud to completely cover the star, which is $\sim$300 days in the case of HMW 15. Coupled with the inferred radius of the star (from its luminosity and effective temperature), we find that the occulting edge crosses the star at the remarkably slow speed of only 50 m s$^{-1}$. This, of course, is the component of the velocity of the cloud that is perpendicular to the (assumed straight) edge that it defines on the star's photosphere. (See Fig. 4 of \citet{h02} for a sketch of how this might appear to an observer.) 
	
	As illustrations of the concept, we consider two specific models of this class, one in which the occulting edge is a circumbinary disk and one in which it is a circumstellar disk. In both cases we assume that the components are a G8 and K4 star with properties based on the eclipsing binary model and that the more massive (bluer) star is eclipsed. We adopt masses of  1 M$_\odot$ and 0.5 M$_\odot$, respectively. For the first example, we assume an orbital period of 4 years (since there is some evidence that the eclipses might recur on that time scale) and a circumbinary disk which may extend to hundreds of AU. To see an eclipse, we must assume that this system is viewed nearly edge-on and that the occulting cloud (the edge of the circumbinary disk, in this scenario) crosses a portion of the orbit of one star. This is easily accomplished if there is a slight angle between the plane of the binary orbit and the top edge of the opaque disk. The required angle of inclination (i) is given by 
	$$tan(i) = {2 R \over v t_{in}} = {0.05 \over v (km\ s^{-1})}$$
where R is the stellar radius, v is the orbital speed of the eclipsed star and t$_{in}$ is the ingress time. For the model under discussion, v is $\sim$7.5 km s$^{-1}$ so i is $\sim$0.4$\degr$. This small angle could easily arise as an inclination of the binary plane to the disk plane.

	Another possible scenario involves eclipse of one component of a binary by the circumstellar disk around its companion. For definiteness, we again consider a 1 M$_\odot$ and a 0.5 M$_\odot$ star, this time separated by 100 AU and having individual circumstellar disks. In this case, the orbital speed of the more massive star is $\sim$1 km s$^{-1}$, so the required inclination angle is $\sim$3$\degr$. Again, this seems quite reasonable dynamically and is not in conflict with observational evidence \citep{jdm}. In this model one would not expect the eclipses to be periodic on any short time scale, but there could be other occultation events, assuming the edge of the disk is not perfectly straight or homogenous. 

	We emphasize that these two particular examples do not exhaust the possibilities. All that is required is a foreground, sharp-edged cloud which is properly oriented with respect to the orbit of the bluer star. Many scenarios are possible to account for the existence of such a cloud. Before exploring this issue in detail, it is imperative to test the basic idea by determining whether HMW 15 is, indeed, a binary star.  Fortunately, it should be rather easy to determine that, since it would need to have components of comparable brightness (out of eclipse). The color and spectral data suggest a G8 + K4 system. High resolution spectroscopy should be able to reveal whether this star is, indeed, a binary system, since the spectrum should appear composite near maximum light. There may also be detectable radial velocity differences between the components, depending on their separation and the phase of the orbit. The experiment should be carried out as soon as possible, while the star is near maximum light, since it is possible that it will begin to fade again almost immediately.
						 
	To summarize, we have detected an apparent eclipse of the G8-K6 PMS star HMW 15 in IC 348 which lasts about 3.5 years, much longer than any other known eclipse. Analysis of the light curve and other data indicates that there is essentially no chance that this is a normal, eclipsing binary star with two PMS components. If the star is single, then it was eclipsed by an optically thin circumstellar or interstellar dust cloud or disk feature. In this case it is hard to understand the flat bottom to the light curve and the color changes, which do not match a normal interstellar reddening law. A currently more attractive hypothesis is that the star is a binary system in which the bluer component was eclipsed by a circumstellar or circumbinary cloud or disk feature. This can be tested by high resolution spectroscopy of the system while it is in its bright state. Along with KH 15D, this object may be a member of small, but important, class of PMS stars whose variability results from eclipses by matter within their circumstellar (or circumbinary) disks.

\acknowledgments

We thank G. Herbig and C. Jordi for kindly providing us with details of their photometry of IC 348, some of it a decade old. We also thank K. Luhman for examining his WFPC images of the field, on which the star is unfortunately saturated. Finally, we thank the many Wesleyan students who contributed to this project by obtaining the data. This research was supported by NASA through its Origins of Solar Systems program.

\clearpage

\clearpage 

\end{document}